# Wireless THz link with optoelectronic transmitter and receiver

T. Harter[1,2]*, S. Ummethala[1,2], M. Blaicher[1,2], S. Muehlbrandt[1,2], S. Wolf[1], M. Weber[1], M. M. H. Adib[1],
J. N. Kemal[1], M. Merboldt[1], F. Boes[3], S. Nellen[4], A. Tessmann[5], M. Walther[5], B. Globisch[4], T. Zwick[3],
W. Freude[1], S. Randel[1], C. Koos[1,2]**

[1]*Institute of Photonics and Quantum Electronics (IPQ), Karlsruhe Institute of Technology (KIT), 76131 Karlsruhe, Germany*
[2]*Institute of Microstructure Technology (IMT), Karlsruhe Institute of Technology (KIT), 76344 Eggenstein-Leopoldshafen, Germany*
[3]*Institute of Radio Frequency Engineering and Electronics (IHE), Karlsruhe Institute of Technology (KIT), 76131 Karlsruhe, Germany*
[4]*Fraunhofer Institute for Telecommunications, Heinrich Hertz Institute (HHI), 10587 Berlin, Germany*
[5]*Fraunhofer Institute for Applied Solid State Physics (IAF), 79108 Freiburg, Germany*
*tobias.harter@kit.edu, **christian.koos@kit.edu

**Photonics might play a key role in future wireless communication systems that operate at THz carrier frequencies. A prime example is the generation of THz data streams by mixing optical signals in high-speed photodetectors. Over the previous years, this concept has enabled a series of wireless transmission experiments at record-high data rates. Reception of THz signals in these experiments, however, still relied on electronic circuits. In this paper, we show that wireless THz receivers can also greatly benefit from optoelectronic signal processing techniques, in particular when carrier frequencies beyond 0.1 THz and wideband tunability over more than an octave is required. Our approach relies on a high-speed photoconductor and a photonic local oscillator for optoelectronic down-conversion of THz data signals to an intermediate frequency band that is easily accessible by conventional microelectronics. By tuning the frequency of the photonic local oscillator, we can cover a wide range of carrier frequencies between 0.03 THz and 0.34 THz. We demonstrate line rates of up to 10 Gbit/s on a single channel and up to 30 Gbit/s on multiple channels over a distance of 58 m. To the best of our knowledge, our experiments represent the first demonstration of a THz transmission link that exploits optoelectronic signal processing techniques both at the transmitter and the receiver.**

## Introduction and background

Data traffic in wireless communication networks is currently doubling every 22 months[1] and will account for more than 60 % of the overall internet traffic by 2021. Sustaining this growth requires advanced network architectures that combine massive deployment of small radio cells[2–4] with powerful backhaul infrastructures, built from high-capacity wireless point-to-point links. Such links may be efficiently realized by exploiting THz carriers in low-loss atmospheric transmission windows[5], thereby offering data rates of tens or even hundreds of Gbit/s. To generate the underlying communication signals at the THz transmitter, optoelectronic signal processing[6–11] has emerged as a particularly promising approach, leading to demonstrations of wireless transmission at line rates of 100 Gbit/s and beyond[12–18]. At the THz receiver, however, the advantages of optoelectronic signal processing have not yet been exploited.

In this paper, we show that wireless THz receivers can benefit from optoelectronic signal processing techniques as well, in particular when carrier frequencies beyond 0.1 THz and wideband tunability are required[19,20]. We exploit a high-speed photoconductor and a photonic local oscillator for terahertz-to-electrical down-conversion over a broad range of frequencies between 0.03 THz and 0.34 THz. In our experiments, we demonstrate a coherent wireless link that operates at a carrier frequency of 0.31 THz and allows line rates of up to 10 Gbit/s on a single channel and up to 30 Gbit/s on multiple channels over a distance of 58 m. To the best of our knowledge, this represents the first demonstration of a THz transmission link that complements optoelectronic signal generation at the transmitter by optoelectronic down-conversion at the receiver.

The vision of a future wireless network architecture is shown in Fig. 1(a). The increasing number of terminal devices and the advent of new data-hungry applications require a dense mesh of small radio cells to provide ubiquitous broadband wireless access[2–4]. Backhauling of these cells relies on high-speed wireless point-to-point links, which are seamlessly integrated into fibre optical networks[21,22]. The high data rates required for wireless backhauling infrastructures are achieved by using carrier frequencies in the range of 0.1 THz to 1 THz (T-waves). Figure 1(b) shows the atmospheric T-wave attenuation as a function of frequency[23], revealing several transmission windows with low attenuation that can be used for wireless communications. For highest flexibility and performance, T-wave transmitters (Tx) and receivers (Rx) should be able to switch between various windows depending on channel occupancy and weather conditions. At the Tx, this can be achieved by optoelectronic T-wave signal generation, Fig. 1(c), which relies on mixing of an optical data signal at a carrier frequency $f_{S,a}$ with continuous-wave (c.w.) tone at frequency $f_{S,b}$ in a high-speed photodiode (optical-to-T-wave, O/T conversion). This leads to a T-wave data signal centred at the carrier frequency $f_S = \left| f_{S,a} - f_{S,b} \right|$, which can be widely tuned by changing the frequency $f_{S,b}$ of the unmodulated optical tone. Note that similar optoelectronic Tx concepts have been used in earlier demonstrations[12–18], but were complemented by an electronic Rx which cannot match the wideband tunability of the Tx. To overcome this limitation, we have implemented an optoelectronic Rx that requires neither an electronically generated local oscillator (LO) nor an electronic mixer for coherent reception, but relies on a photoconductor that is driven by a photonic LO instead, Fig. 1d. The photonic LO is generated by superimposing two optical c.w. tones with frequencies $f_{LO,a}$ and $f_{LO,b}$ and is coupled to the photoconductor[24–26] for down-conversion of the T-wave data signal to an intermediate frequency band that is easily accessible by conventional microelectronics (T-wave-to-electrical, T/E conversion).

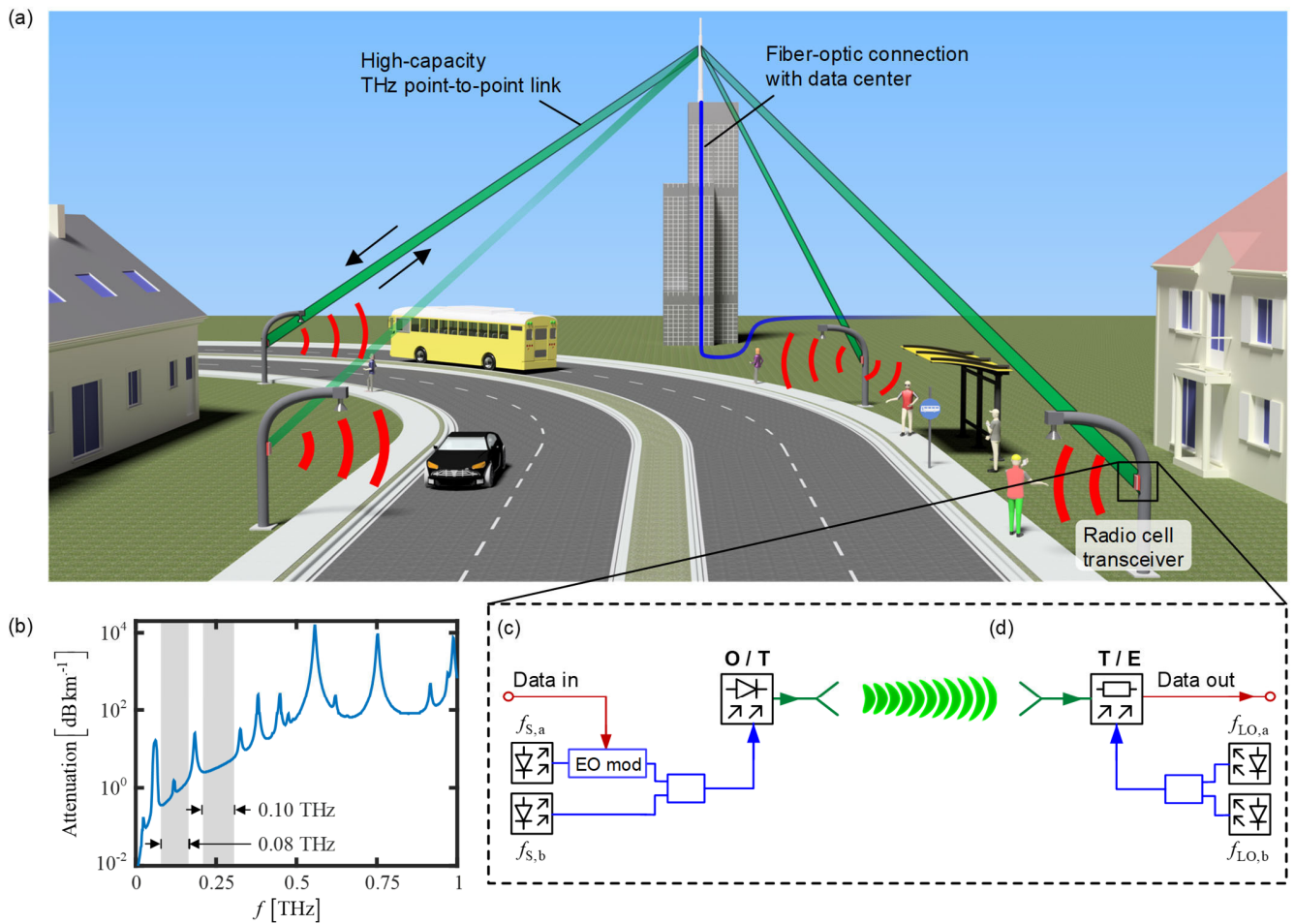

**Fig. 1.** T-wave wireless infrastructure using optoelectronic techniques. **(a)** Vision of a future wireless network architecture. A dense mesh of small radio cells provides broadband wireless access to a vast number of users and devices. The high data rates required for the underlying wireless backhauling infrastructures are provided by high-speed wireless point-to-point links that are operated at THz frequencies and that can be efficiently interfaced with fibre-optic networks. **(b)** T-wave atmospheric attenuation for standard conditions[23] (temperature of 15 °C, water-vapour content of 7.5 g/m$^3$). Various windows with low attenuation can be used for T-wave communications. Our Rx allows operation over a wide range of frequencies between 0.03 THz and 0.34 THz, in which the atmospheric attenuation is small enough to permit transmission over technically relevant distances. **(c)** Optoelectronic T-wave signal generation. The data signal is modulated on an optical continuous-wave (c.w.) tone with frequency $f_{S,a}$ by an electro-optic modulator (EO mod). The modulated signal is superimposed with an unmodulated c.w. tone $f_{S,b}$. The optical signal is converted to a T-wave signal in a high-speed photodiode (optical-to-T-wave conversion, O/T) which is radiated into free space by an antenna. The carrier frequency of the T-wave is given by the frequency difference $f_S = |f_{S,a} - f_{S,b}|$. **(d)** Optoelectronic coherent T-wave reception. The T-wave data signal is down-converted in a photoconductor where the optical power beat $f_{LO} = |f_{LO,a} - f_{LO,b}|$ of two unmodulated c.w. tones acts as photonic local oscillator (T-wave-to-electric conversion, T/E).

Note that optically driven photoconductors have previously been used for down-conversion of THz waves in spectroscopy systems[24–28]. These demonstrations, however, are usually based on THz Tx and Rx that are driven by a common pair of lasers ($f_{S,a} = f_{LO,a}$, $f_{S,b} = f_{LO,b}$) for homodyne reception, and they rely on narrowband detection schemes with typical averaging times of the order of 1 ms and bandwidths of a few kHz for highly sensitive acquisition of small power levels. In our work, we advance these concepts to enable wireless data transmission at GHz bandwidth over technically relevant distances, using the complex amplitude of the THz wave for encoding of information. To this end, we make use of advanced photoconductors with engineered carrier lifetime[24], we combine them with high-speed transimpedance amplifiers [29] into a perfectly shielded read-out circuit, and we exploit heterodyne detection in combination with advanced digital signal

3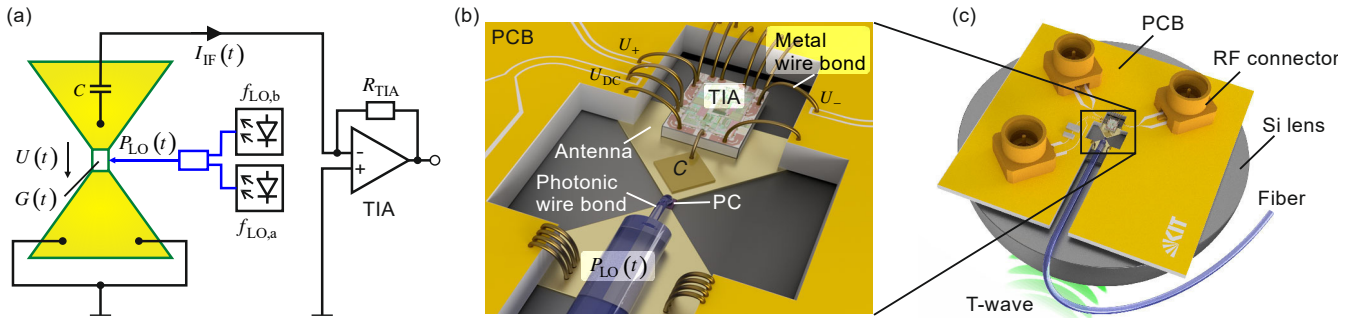

**Fig. 2.** Optoelectronic coherent T-wave Rx. **(a)** Schematic of the Rx module. The T-wave signal is received by a bow-tie antenna with a photo-conductor between the antenna feed points. This leads to a T-wave voltage signal $U(t)$ applied to the photoconductor, centered around a carrier frequency $f_S$. At the same time the photoconductance $G(t)$ is modulated at a frequency $f_{LO}$ by the power beat of two unmodulated laser tones, $f_{LO} = |f_{LO,a} - f_{LO,b}|$. Both effects combined lead to a down-converted current $I_{IF}(t)$ oscillating at the difference frequency $|f_S - f_{LO}|$. The current is amplified by a transimpedance amplifier (TIA) having a transimpedance $R_{TIA}$. The capacitor $C = 1\,\text{nF}$ blocks direct currents at the input circuit of the TIA. **(b)** Schematic of Rx module. The photoconductor (PC) and the antenna is electrically bonded to the TIA. The differential outputs of the TIA ($U_+$ and $U_-$) are connected to a printed circuit board (PCB, a gold-plated alumina ceramic substrate) which includes RF connectors through which the signals are fed to high-speed oscilloscopes for further analysis. The photoconductor is illuminated with the optical power $P_{LO}(t)$ by a fiber and a 3D-printed photonic wire bond. **(c)** The Rx assembly of photoconductor, antenna, TIA, and PCB is glued on a silicon lens for efficient coupling of the incoming T-wave to the antenna.

processing to overcome phase noise and drift associated with the free-running photonic LO at the Rx. In the following, we give details about the optoelectronic Rx module and the demonstration of the T-wave link.

### Implementation of optoelectronic receiver

The concept and the implementation of the optoelectronic Rx is illustrated in Fig. 2(a). The T-wave data signal, oscillating at an angular carrier frequency $\omega_S = 2\pi f_S$, is received by a bow-tie antenna, resulting in a T-wave voltage $U(t)$ across the feed points,

$$U(t) = \hat{U}_S(t)\cos(\omega_S t + \varphi_S(t)). \quad (1)$$

In this relation, $\hat{U}_S(t)$ is the modulated T-wave voltage amplitude, and $\varphi_S(t)$ is the associated modulated phase. The antenna feed points are connected to the photoconductor $G$, which is illuminated by the photonic LO that provides a time-dependent optical power, which oscillates at a frequency $\omega_{LO} = 2\pi |f_{LO,a} - f_{LO,b}|$ and has an amplitude $\hat{P}_{LO,1}$,

$$P_{LO}(t) = P_{LO,0} + \hat{P}_{LO,1}\cos(\omega_{LO} t + \varphi_{LO}). \quad (2)$$

The free carriers generated by the absorbed optical power change the photoconductance according to

$$G(t) = \text{G}\,P_{LO}(t) = G_0 + \hat{G}_{LO}\cos(\omega_{LO} t + \varphi_{LO}), \quad (3)$$

where G denotes a proportionality constant that describes the sensitivity of the photoconductor. The resulting current $I(t)$ through the photoconductor is given by the product of the time-varying conductance $G(t)$ and the T-wave voltage $U(t)$, leading to mixing of the T-wave signal centred around $\omega_S$ and the optical LO power oscillating at $\omega_{LO}$. After processing of the current by subsequent electronics such as a transimpedance amplifier (TIA), only the low-frequency-components of the mixing product remain, leading to a down-converted current at an intermediate frequency $\omega_{IF} = |\omega_S - \omega_{LO}|$,

$$I_{IF}(t) = \tfrac{1}{2}\hat{G}_{LO}\hat{U}_S(t)\cos(\omega_{IF} t + \varphi_S(t) - \varphi_{LO}). \quad (4)$$

This intermediate signal contains the amplitude and phase information of the T-wave data signal. A more detailed derivation of Eqs. (1)-(4) can be found in Supplementary Section 1.

Figure 2(b) illustrates the technical implementation of the Rx module used for our experiments. The photoconductor[24] (PC) is connected to a bow-tie antenna which is electrically coupled to a TIA by a metal wire bond. The photoconductor is operated without any DC bias, and a decoupling capacitor $C = 1\,\text{nF}$ is used to isolate the device from the bias that is effective at the input of the TIA, see Supplementary Section 2. The output of the TIA is electrically connected to a printed circuit board. The photoconductor is illuminated from the top with the optical power $P_{LO}(t)$, which is coupled from the horizontally positioned fibre by a photonic wire bond[30]. The entire assembly is glued on a silicon lens which focuses the incoming T-wave to the antenna, see Fig. 2(c), and the assembly is placed in metal housing for effective electromagnetic shielding. A microscope image of the fabricated device and a more detailed description of the Rx module in terms of conversion efficiency, bandwidth and noise can be found in Supplementary Section 2.

### Demonstration of wireless THz links

In the following we demonstrate the viability of our receiver concept in a series of experiments, covering both single-channel and multi-channel transmission of THz data signals. For the single-channel experiments, our focus was on demonstrating the wideband tunability of the carrier frequency. The multi-channel experiment shows the scalability of the approach towards high-throughput parallel transmission.

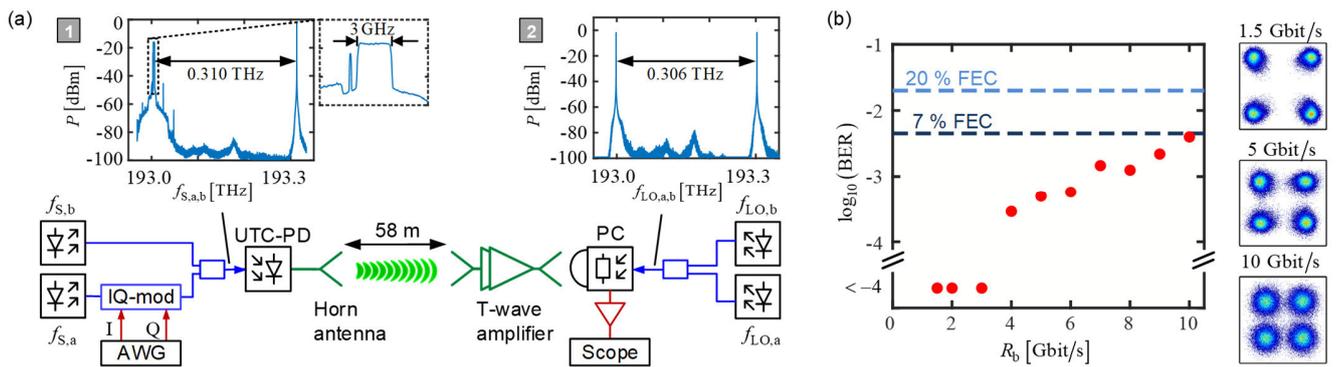

**Fig. 3.** Experimental demonstration of T-wave wireless transmission. **(a)** At the Tx, an optical quadrature phase shift keying (QPSK) signal at a carrier frequency $f_{S,a}$ is generated by an IQ-modulator and an arbitrary-waveform generator (AWG). The optical signal is then superimposed with an unmodulated optical c.w. tone $f_{S,b}$ and converted to a T-wave data signal by a high-speed uni-travelling carrier photodiode (UTC-PD). The T-wave is radiated into free space by a horn antenna. The frequency of the T-wave carrier depends on the frequency difference of the lasers $f_S = |f_{S,a} - f_{S,b}| = 0.310\,\text{THz}$. At the Rx, two c.w. laser tones with frequencies $f_{LO,a}$ and $f_{LO,b}$ are superimposed to generate an optical power beat, which acts as local oscillator ("photonic LO") for coherent down-conversion of the T-wave by an antenna-coupled photoconductor (PC). The wireless transmission link spans a distance of 58 m. To compensate the transmission loss, we use a two-stage T-wave amplifier in front of the Rx. The received data signal is recorded by a real-time oscilloscope and offline digital signal processing is used to analyze the data. **Inset 1:** Optical spectrum (180 MHz resolution bandwidth, RBW) at the Tx for a 3 GBd QPSK data stream and a T-wave carrier frequency of $f_S = 0.310\,\text{THz}$. **Inset 2:** Optical spectrum (180 MHz RBW) of the photonic LO at the Rx for $f_{LO} = 0.306\,\text{THz}$. **(b)** Measured bit error ratio (BER) for QPSK data streams with various line rates $R_b$ transmitted at a T-wave carrier frequency of around 0.310 THz. For line rates up to 10 Gbit/s a BER below the threshold for forward-error correction (FEC) with 7 % overhead is achieved.

### Single-channel transmission and wideband tunability

The wireless transmission system for single-channel transmission is illustrated in Fig. 3(a). As data source we use an arbitrary-waveform generator (AWG) which provides a quadrature phase shift keying (QPSK) data to an optical IQ-modulator. The modulator is fed by a tunable laser with an optical carrier frequency $f_{S,a}$. The modulated data signal is superimposed by an unmodulated optical carrier with frequency $f_{S,b}$. The optical power spectrum of the data signal and the unmodulated laser tone for a 3GBd QPSK signal and a frequency spacing of $f_S = |f_{S,a} - f_{S,b}| = 0.310\,\text{THz}$ is shown in Inset 1 of Fig. 3(a). For O/T-conversion, we use a commercially available uni-travelling-carrier photodiode[31] (UTC-PD). The T-wave signal is then radiated into free space by a horn antenna and transmitted over a distance of 58 m, limited only by the size of our building. At the Rx, we use a horn antenna to couple the T-wave signal into a WR 3.4 hollow waveguide that is connected to a two-stage T-wave amplifier[32] which compensates the transmission loss. At the output of the amplifier another horn antenna is used to feed the signal to the silicon lens of the T-wave Rx. T/E-conversion is then performed in the photoconductor which is illuminated by a photonic LO, see power spectrum in Inset 2 of Fig. 3(a). A comprehensive description of the transmission setup and a characterization of the T-wave amplifiers and the UTC-PD is given in Supplementary Sections 3 and 4.

After down-conversion, the electrical signals are recorded by a real-time oscilloscope and stored for offline digital signal processing (DSP). The T-wave Rx relies on a heterodyne detection scheme $(f_S \neq f_{LO})$, where the photonic LO is placed at the edge of the T-wave data spectrum, which leads to electrical signals centred around the intermediate frequency $f_{IF} = |f_S - f_{LO}|$. The in-phase and the quadrature components of the QPSK baseband signals are then extracted from the intermediate signals by DSP, comprising standard procedures such as digital frequency down-conversion, timing recovery, constant-modulus equalization, frequency offset compensation, and carrier phase estimation.

Figure 3(b) shows the bit error ratio (BER) measured for various line rates $R_b$ at a carrier frequency of 0.31 THz. For line rates below 3 Gbit/s, no errors are measured in our recording length of $10^5$ symbols, demonstrating the excellent performance of the optoelectronic Rx. The constellation diagrams for line rates of 1.5 Gbit/s, 5 Gbit/s and 10 Gbit/s are shown in the insets. For larger line rates, the received signal quality decreases mainly due to limitations of the TIA in the intermediate-frequency circuit. The TIA has a specified bandwidth of only 1.4 GHz and larger line rates would require a more broadband device. With the current TIA, we could transmit 10 Gbit/s with a BER below the threshold of forward-error correction (FEC) with 7 % overhead. Note that the transmission distance of 58 m was dictated by space limitations. With the current components, it would be possible to bridge roughly twice the distance by doubling the optical power at the Tx.

To the best of our knowledge, our experiments represent the first demonstration of a THz transmission link that complements optoelectronic generation of T-wave signals at the Tx by opto-electronic down-conversion at the Rx. Similar schemes have also been demonstrated[33,34] at lower carrier frequencies, relying on a UTC-PD for optoelectronic T/E conversion. Using this Rx, a line rate of 5 Gbit/s at a carrier frequency of 35.1 GHz and a line rate of 1 Gbit/s at a carrier frequency of 60 GHz have been demonstrated with transmission distances of 1.3 m and 0.55 m, respectively. Our work relies on photoconductors with excellent linearity both with respect to the voltage and the applied optical power[19,20] and clearly demonstrates the vast potential of optoelectronic down-conversion for T-wave communications at tens of Gbit/s over extended distances.

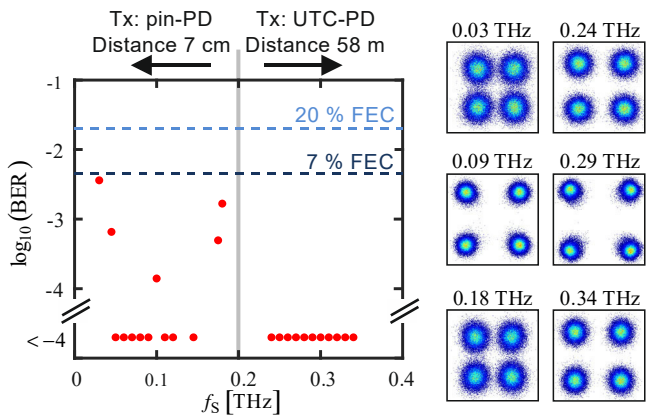

**Fig. 4.** BER and constellation diagrams for various carrier frequencies at a line rate of $R_b = 2$ Gbit/s. For the Tx we used two different types of photodiodes depending on the T-wave frequency. For T-wave frequencies $0.03\,\text{THz} \leq f_S \leq 0.18\,\text{THz}$ we use a lens-coupled pin photodiode (pin-PD). For simplicity, the transmission experiments in this frequency range were performed over a reduced transmission distance of 7 cm only, which could be bridged without any amplifiers. For T-wave frequencies $0.24\,\text{THz} \leq f_S \leq 0.34\,\text{THz}$, a waveguide-coupled uni-travelling-carrier photodiode (UTC-PD) is used in combination with a cascade of two T-wave amplifiers, as described in Fig. 3. For both cases, the same Rx module as in Fig. 2 was used, demonstrating its large tunability.

To further demonstrate the flexibility of the optoelectronic Rx, we transmit 2 Gbit/s data streams at various carrier frequencies covering the entire range between 0.03 THz and 0.34 THz. Note that the setup shown in Fig. 3(a) only allows to cover the frequency range between 0.24 THz and 0.34 THz due to bandwidth limitations of both the UTC-PD and the T-wave amplifiers. For transmission at frequencies between 0.03 THz and 0.18 THz, we omitted the amplifiers and replaced the UTC-PD by a pin-PD. The measured BER and some exemplary constellation diagrams of the transmission experiments are shown in Fig. 4. For carrier frequencies between 0.24 THz and 0.34 THz, no errors were measured in our recordings such that we can only specify an upper limit of $10^{-4}$ for the BER. For carrier frequencies between 0.03 THz and 0.18 THz, comparable performance was obtained. For simplicity, the transmission experiments in the lower frequency range were performed over a transmission distance of 7 cm only, which could be bridged without any amplifiers. The range could be easily extended to tens or hundreds of meters by using Rx antennae that are optimized for lower frequencies in combination with amplifiers. Note also that the data points between 0.03 THz and 0.18 THz were taken on a slightly irregular frequency grid, thereby avoiding some carrier frequencies for which the free-space link of our setup features low transmission. This does not represent a fundamental problem, but was caused by fading due to uncontrolled reflections in the beam path. The associated power variations could be overcome by using amplifiers with adaptive gain or by optimizing the beam path for each frequency point individually. This fading in combination with frequency-dependent Tx power and a decreasing gain of the on-chip bow-tie antenna for low frequencies is also the reason for the degraded performance of the transmission experiments at carrier frequencies of 0.03 THz, 0.04 THz, 0.10 THz, and 0.18 THz. Still, these experiments demonstrate that the same receiver concept as in Fig. 2 can be used for carrier frequencies in a range of $0.03\,\text{THz} \leq f_S \leq 0.340\,\text{THz}$, i.e. over more than a decade.

**Multi-channel transmission**

We also investigate the receiver in a multicarrier transmission experiment at carrier frequencies between 0.287 THz and 0.325 THz. We simultaneously transmit up to 20 T-wave channels (Ch 1 … 20) spaced by 2 GHz, where each channel is operated with a QPSK signal at a symbol rate (line rate) of 0.75 GBd (1.5 Gbit/s). To keep the experimental setup simple, we use a single broadband AWG and a single IQ-modulator to generate an optical signal that simultaneously contains all channels, which is then converted to the THz range by a UTC-PD. This approach allows us to re-use the experimental setup shown on Fig. 3(a). Alternatively, multiple optical carriers and less broadband devices in combination with optical multiplexing could have been used to generate the optical channels[14]. Figure 5(a) shows the optical spectrum containing all 20 channels, measured after the beam combiner before the UTC-PD. For compensating the spectral roll-off of the UTC-PD and the T-wave amplifiers, the channels at the edges of the T-wave transmission band are pre-emphasized, see inset of Fig. 5(a). Figure 5(b) shows the spectrum of the two optical lines which are used as photonic LO for reception of Ch 20, which features the highest THz carrier frequency of $f_S = 0.325\,\text{THz}$. Note that we again used heterodyne detection at the receiver and hence chose a T-wave LO frequency $f_{LO} = 0.326\,\text{THz}$ close to the spectral edge of Ch 20. It is also worth mentioning that the 0.75 GBd T-wave channels were transmitted on a 2 GHz grid to avoid interference of data signals from neighbouring channels after down-conversion to the intermediate frequency band. This leads to un-used spectral regions of approximately 1.2 GHz between the T-wave channels, which could be avoided by optoelectronic down-conversion schemes that allow simultaneous extraction of the in-phase and the quadrature component of the T-wave signal. Figure 5(c) shows the BER for transmission experiments with 6, 12, and 20 channels. For transmission of 12 channels (aggregate line rate 18 Gbit/s), the BER stays below the 7% FEC limit, whereas for 20 channels (30 Gbit/s), 20 % FEC overhead is required. In Fig. 5(d), the constellation diagrams of the 20 channel experiment are displayed. Note that in our current experiments, the data per channel was only limited by the bandwidth of the TIA. Using more broadband devices[35], symbol rates of more than 25 GBd and data rates of more than

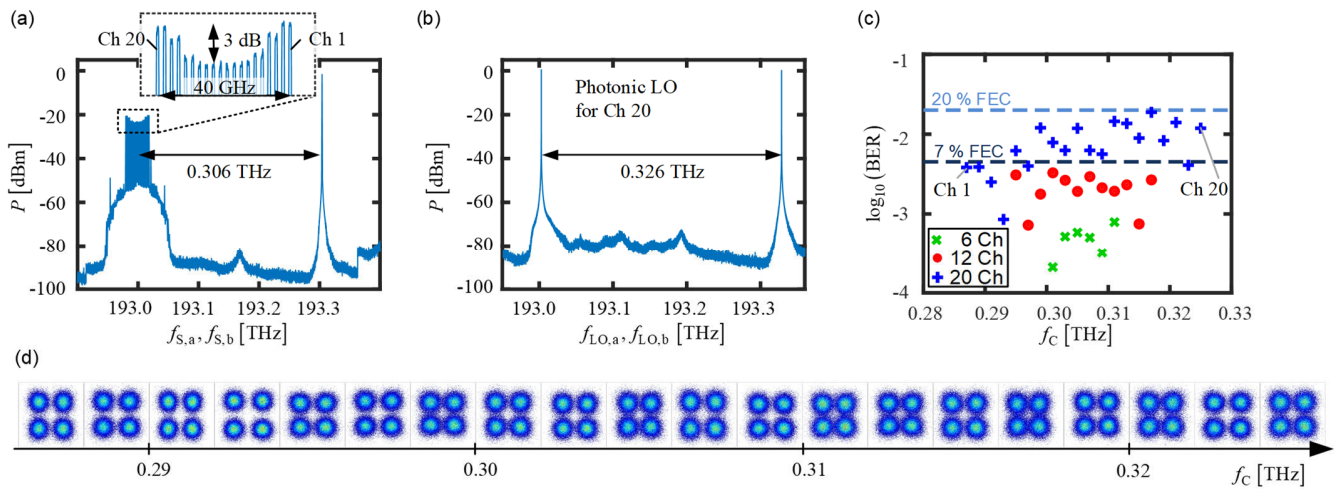

**Fig. 5.** Multi-channel T-wave transmission. **(a)** Optical spectrum (180 MHz RBW) at the Tx for a signal containing 20 channels. Each channel is modulated with pulses having a raised-cosine spectrum with a roll-off factor of 0.1 and carries a 0.75 GBd QPSK signal. The channels are spaced by 2 GHz. In the UTC-PD, the channels are simultaneously down-converted to a T-wave frequency band centred at 0.306 THz. **(b)** Optical spectra (180 MHz RBW) at the receiver for detection of Ch 20. For heterodyne detection, the T-wave LO frequency $f_{LO} = 0.326\,\text{THz}$ is chosen close to the spectral edge of the channel. **(c)** Measured BER for various numbers of channels. For 12 (20) channels, the BER is below the 7% (20%) threshold for forward error correction (FEC). This corresponds to an aggregate line rate of 18 Gbit/s (30 Gbit/s). **(d)** Constellation diagrams for all 20 channels offering an aggregate line rate of 30 Gbit/s.

50 Gbits/s per channel may be achieved in the future. Note also that in our multi-channel experiments the different channels were measured sequentially. While this is a usual approach in multi-channel THz transmission experiments[13,14,18], parallel reception of the entire data stream would be desirable. In this context, T-wave demultiplexers might become highly relevant in the future[36,37].

## Summary

In summary, we showed a first demonstration of a coherent wireless THz communication system using optoelectronic signal processing both at the transmitter and at the receiver. Our experiments show that the same receiver concept can be used over a broad frequency range $0.03\,\text{THz} \leq f_S \leq 0.340\,\text{THz}$, spanning more than a decade. We transmit a line rate of 10 Gbit/s using a single T-wave channel at a carrier frequency of 0.31 THz with a BER below the 7% FEC limit. In this experiment, the line rate was limited by the bandwidth of the transimpedance amplifier, but not by the transmitter and receiver scheme. We further demonstrate multi-channel transmission using up to 20 carriers with frequencies in the range between 0.287 THz and 0.325 THz. This leads to an aggregate line rate of 30 Gbit/s with a BER below the threshold for a FEC with 20 % overhead. The single and the multi-channel T-wave link bridges a distance of 58 m. In the future, optoelectronic T-wave receivers may exploit integrated frequency converters that may be efficiently realized on the silicon plasmonic platform[9]. Our findings demonstrate that coherent T-wave receivers with an optoelectronic, widely tunable local oscillator may build the base for a novel class of THz communication systems.

**Funding**

ERC Consolidator Grant (TeraSHAPE, 773248); Alfried Krupp von Bohlen und Halbach Foundation; Helmholtz International Research School of Teratronics (HIRST); Karlsruhe School of Optics and Photonics (KSOP); Karlsruhe Nano Micro Facility (KNMF).

**Acknowledgment.**
The authors thank NTT Electronics (NEL) for providing the UTC-PD for this experiment. Packaging and assembly of the receiver module was supported by Andreas Lipp and Oswald Speck

# Wireless THz link with optoelectronic transmitter and receiver (Supplementary Information)

T. Harter[1,2,*], S. Ummethala[1,2], M. Blaicher[1,2], S. Muehlbrandt[1,2], S. Wolf[1], M. Weber[1], M. M. H. Adib[1],
J. N. Kemal[1], M. Merboldt[1], F. Boes[3], S. Nellen[4], A. Tessmann[5], M. Walther[5], B. Globisch[4], T. Zwick[3],
W. Freude[1], S. Randel[1], C. Koos[1,2,**]

[1]*Institute of Photonics and Quantum Electronics (IPQ), Karlsruhe Institute of Technology (KIT), 76131 Karlsruhe, Germany*
[2]*Institute of Microstructure Technology (IMT), Karlsruhe Institute of Technology (KIT), 76344 Eggenstein-Leopoldshafen, Germany*
[3]*Institute of Radio Frequency Engineering and Electronics (IHE), Karlsruhe Institute of Technology (KIT), 76131 Karlsruhe, Germany*
[4]*Fraunhofer Institute for Telecommunications, Heinrich Hertz Institute (HHI), 10587 Berlin, Germany*
[5]*Fraunhofer Institute for Applied Solid State Physics (IAF), 79108 Freiburg, Germany*
[*]tobias.harter@kit.edu, [**]christian.koos@kit.edu

## 1. Detailed derivation of formulae

In the main paper, we show coherent wireless THz communication using an optoelectronic receiver[1–3] and a tunable photonic local oscillator (LO). The concept of optoelectronic down-conversion in a photoconductive T-wave receiver (Rx) is illustrated in Fig. 2(a) of the main paper. In the following, we give a detailed derivation of the associated formulae.

The T-wave data signal from the transmitter (Tx) at an angular carrier frequency $\omega_S = 2\pi f_S$ is received by a bow-tie antenna resulting in a T-wave voltage $U(t)$ across its feed points,

$$U(t) = \hat{U}_S(t)\cos(\omega_S t + \varphi_S(t)). \quad \text{(S1)}$$

In this relation, $\hat{U}_S(t)$ is the modulated T-wave voltage amplitude, and $\varphi_S(t)$ is the associated modulated phase. The antenna feed points are connected to a photoconductor $G$, which is illuminated by the superposition of two unmodulated optical fields $E_{LO,a}(t)$ and $E_{LO,b}(t)$ with frequencies $\omega_{LO,a}$, $\omega_{LO,b}$ and amplitudes $\hat{E}_{LO,a}$, $\hat{E}_{LO,b}$,

$$\begin{aligned}E_{LO,a}(t) &= \hat{E}_{LO,a}\cos(\omega_{LO,a}t),\\ E_{LO,b}(t) &= \hat{E}_{LO,b}\cos(\omega_{LO,b}t).\end{aligned} \quad \text{(S2)}$$

This leads to an optical power, which oscillates at a frequency $\omega_{LO} = |\omega_{LO,a} - \omega_{LO,b}|$ and has an amplitude $\hat{P}_{LO,1}$,

$$P_{LO}(t) = P_{LO,0} + \hat{P}_{LO,1}\cos(\omega_{LO}t + \varphi_{LO}). \quad \text{(S3)}$$

The quantities $P_{LO,0}$, $\hat{P}_{LO,1}$ are expressed by the (normalized) electrical field strengths

$$P_{LO,0} = \tfrac{1}{2}(\hat{E}_{LO,a}^2 + \hat{E}_{LO,b}^2), \quad \hat{P}_{LO,1} = \hat{E}_{LO,a}\hat{E}_{LO,b}. \quad \text{(S4)}$$

The photocarriers generated by the absorbed optical power change the photoconductance according to

$$G(t) = \mathrm{G}\, P_{LO}(t) = G_0 + \hat{G}_1\cos(\omega_{LO}t + \varphi_{LO}), \quad \text{(S5)}$$

where G denotes a proportionality constant that describes the sensitivity of the photoconductor. The resulting current $I(t)$ through the photoconductor is then given by the product of the time-varying conductance $G(t)$ and the time-varying voltage,

$$\begin{aligned}I(t) &= G(t)U(t)\\ &= \underbrace{G_0\hat{U}_S(t)\cos(\omega_S t + \varphi_S(t))}_{(1)}\\ &\quad + \underbrace{\tfrac{1}{2}\hat{G}_{LO}\hat{U}_S(t)\cos((\omega_S + \omega_{LO})t + \varphi_S(t) + \varphi_{LO})}_{(2)}\\ &\quad + \underbrace{\tfrac{1}{2}\hat{G}_{LO}\hat{U}_S(t)\cos((\omega_S - \omega_{LO})t + \varphi_S(t) - \varphi_{LO})}_{(3)}.\end{aligned} \quad \text{(S6)}$$

After amplification of the current $I(t)$ using a transimpedance amplifier (TIA), only the low-frequency-part remains and expressions (1) and (2) of Eq. (S6) are filtered out. This leads to a down-converted current at the intermediate frequency $\omega_{IF} = |\omega_S - \omega_{LO}|$,

$$I_{IF}(t) = \tfrac{1}{2}\hat{G}_{LO}\hat{U}_S(t)\cos(\omega_{IF}t + \varphi_S(t) - \varphi_{LO}). \quad \text{(S7)}$$

The intermediate signal contains the amplitude and phase information of the T-wave data signal and can be processed by low-frequency electronics.

## 2. T-wave receiver

This section gives details of the implementation and the characterization of the optoelectronic Rx used in our experiments. Figure S1 shows images of our Rx module. The photoconductor[2] (PC) is in direct contact with a bow-tie antenna, see Fig. S1(a). The antenna is electrically bonded to a transimpedance amplifier (TIA, Maxim Integrated[4] PHY1097) for processing the down-converted intermediate-frequency current. The TIA is designed for amplification of receiver signals in a passive optical network, where the photodiode is reversed biased by the TIA. In our application, the photoconductor does not require a bias voltage and hence we use a capacitor $C$ to decouple the photoconductor from the bias at the TIA input terminals. The output of the TIA is electrically bonded to a printed circuit board (PCB) having a gold-plated alumina ceramic substrate. The photoconductor is illuminated from the top with the time-dependent optical power $P_{\text{LO}}(t)$, which is coupled to the active region of the device from the horizontally positioned fiber by a photonic wire bond[5,6], see inset of Fig. S1(a). The assembly is placed on a silicon lens which focuses the incoming T-wave onto the antenna, see Fig. 2(c) of the main manuscript. All components are placed in a metal housing for electromagnetic shielding of the Rx circuits and for simplified handing of the Rx. The fully packaged Rx is shown in Fig. S2(b). The photonic LO is fed to the Rx with a fiber, and the down-converted RF data signal ("data out") is processed further by standard laboratory equipment.

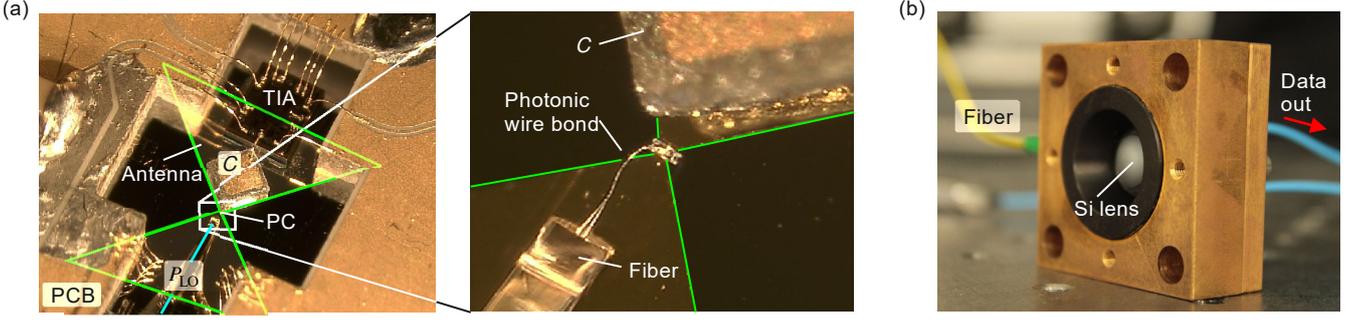

**Fig. S1.** Optoelectronic coherent T-wave Rx. **(a)** Microscope image of the Rx module. The photoconductor (PC) is connected to an on-chip bow-tie antenna, which is electrically bonded to a transimpedance amplifier (TIA). The capacitor $C$ = 1 nF isolates the PC from the DC bias voltage that is supplied by the TIA at its input. The output of the TIA is electrically connected to a printed circuit board (PCB) featuring a gold-plated alumina ceramic substrate. The PCB includes sub-miniature plugs for off-chip RF and DC connections. The photoconductor is illuminated with the time-dependent optical power $P_{\text{LO}}(t)$ by a fiber and a 3D-printed photonic wire bond, see inset for details. **(b)** Fully packaged Rx module. The photoconductor, antenna, TIA, and PCB are glued to a silicon lens for coupling the T-wave to the on-chip antenna. All components are placed inside a metal housing for electromagnetic shielding of the Rx circuits. The photonic LO is fed to the Rx with a fiber. The down-converted data signal ("data out") is processed further by standard RF-equipment connected to the sub-miniature plugs on the PCB.

In the following, we give a detailed characterization of the Rx in terms of conversion efficiency, bandwidth, and noise.

**Conversion efficiency**

First we quantify the frequency-dependent response of the photoconductor connected to a bow-tie antenna. We define the conversion efficiency $\eta$ as the ratio of the output power at the intermediate frequency in a $50\,\Omega$ load resistor related to the incident THz power $P_{\text{THz}}$. The conversion efficiency is measured with a photoconductor very similar to the one used for the data transmission experiment. Details are published elsewhere[1].

For this measurement, the same continuous-wave (c.w.) lasers are used for both the Tx and the Rx LO, i. e, $\omega_{\text{S}} = \omega_{\text{LO}}$. One of the Tx lasers is phase modulated[7,8] by a ramp with maximum amplitude $2\pi$ and repetition period $2\pi/\omega_{\text{m}}$. The resulting voltage at the Rx antenna feed points is modulated according to Eq. (S1), $U(t) = \hat{U}_{\text{S}} \cos\left((\omega_{\text{S}} + \omega_{\text{m}})t - \varphi_{\text{S}}\right)$. After down-conversion, the current according to Eq. (S7) becomes $I_{\text{IF}}(t) = \hat{I}_{\text{IF}} \cos(\omega_{\text{m}} t + \varphi_{\text{S}} - \varphi_{\text{LO}})$ with $\hat{I}_{\text{IF}} = \frac{1}{2}\hat{G}_{\text{LO}}\hat{U}_{\text{S}}$. The current amplitude $\hat{I}_{\text{IF}}$ is measured with a lock-in amplifier (LIA) tuned to the repetition frequency $\omega_{\text{m}}$. We change the T-wave frequency $\omega_{\text{S}} = \omega_{\text{LO}}$ and measure $\hat{I}_{\text{IF}}$ with the LIA along with the incident THz power $P_{\text{THz}}$ using a calibrated pyroelectric thin film detector (Sensor- und Lasertechnik GmbH, THz20). For the conversion efficiency $\eta$ we define

$$\eta = \frac{\frac{1}{2}\hat{I}_{\text{IF}}^2 \times 50\,\Omega}{P_{\text{THZ}}}. \quad \text{(S8)}$$

Figure S2 shows the conversion efficiency of the photoconductive Rx in dependence of the frequency $f_{\text{S}}$. The grey hatched area indicates the frequency range used in our experiments. In general, the T-wave bandwidth of a photoconductor is limited by the lifetime $\tau$ of the free carriers that are generated by the incident optical signal. This lifetime can be reduced by low-temperature growth of the associated III-V materials[2]. The frequency response of our device shows a roll-off of the conversion efficiency for frequencies beyond $f_{\text{S},\tau} = 0.2\,\text{THz}$, corresponding to a carrier lifetime of $\tau = 1/(2\pi f_{\text{S},\tau}) = 0.8\,\text{ps}$. The conversion efficiency drops by 10 dB, but remains virtually constant in a frequency range of $0.3\,\text{THz} \leq f_{\text{S}} \leq 0.6\,\text{THz}$.

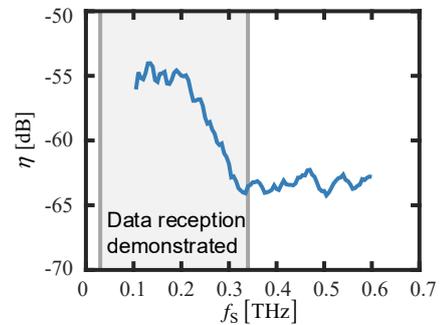

**Fig. S2.** Conversion efficiency of the photoconductive Rx. The gray area marks the frequency range that is used for the data transmission experiments. The conversion efficiency for frequencies larger than 0.34 THz is reduced by 10 dB, but remains constant up to at least 0.6 THz.

## Bandwidth

While the optoelectronic part of the T-wave Rx is extremely broadband, the maximum received data rate is limited by the TIA (Maxim Integrated, PHY1097 [4]) which features a 3 dB bandwidth of 1.4 GHz and is used for amplifying the down-converted signal $I_{IF}(t)$. The bandwidth is specified for a standard application where the TIA is used in combination with a photodetector in a passive optical network. To measure the overall module bandwidth, we use the setup shown in Fig S3(a). A c.w. T-wave tone at a constant frequency $f_S = |f_{S,a} - f_{S,b}| = 0.1\,\text{THz}$ is generated. At the Rx, the T-wave tone is down-converted using a photonic local oscillator with frequency $f_{LO} = |f_{LO,a} - f_{LO,b}| \neq f_S$, Eq. (7). We then measure the power $\frac{1}{2}\hat{I}_{IF}^2 R_{TIA}$ of the down-converted IF tone at frequency $f_{IF} = |f_S - f_{LO}|$ using an electrical spectrum analyzer (ESA). By tuning the photonic local-oscillator frequency $f_{LO}$, the IF frequency response of the Rx module is obtained. The measured IF frequency response is normalized to its value in the frequency range between 100 MHz and 200 MHz, and the result is shown in Fig. S3(b). The Rx response is fairly constant up to 0.75 GHz and drops quickly for higher frequencies. The measured bandwidth is smaller than expected from the TIA specifications. We attribute this to fact that the impedance of the PC at the TIA input differs from that of a high-speed photodiode, for which the device is optimized.

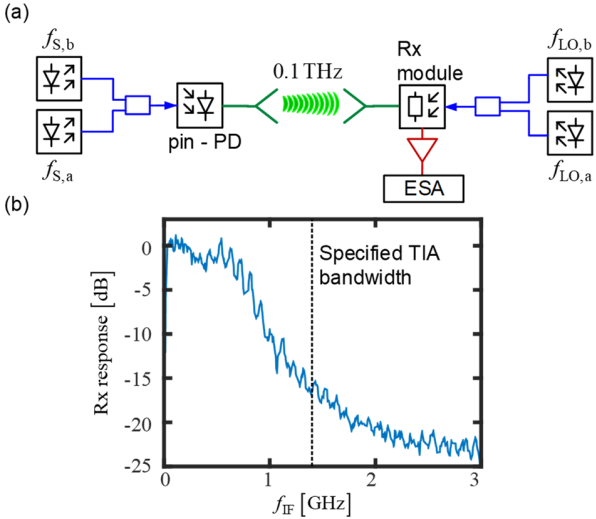

**Fig. S3.** IF frequency response of the Rx module. **(a)** Measurement setup. A continuous-wave (c.w.) T-wave tone at a constant frequency of $f_S = |f_{S,a} - f_{S,b}| = 0.1\,\text{THz}$ is generated by photomixing of two optical c.w. laser tones. At the Rx, the T-wave signal is down-converted using a photonic local oscillator with frequency $f_{LO} = |f_{LO,a} - f_{LO,b}|$, which differs from $f_S$ by the targeted intermediate frequency $f_{IF} = |f_S - f_{LO}|$. The power of the down-converted c.w. tone is measured by an electrical spectrum analyzer (ESA). By tuning the local oscillator frequency $f_{LO}$, we obtain the IF frequency response of the Rx. **(b)** Measured IF frequency response of the Rx module.

## Noise

Figure S4 shows the noise characteristic of the packaged Rx module measured without an incident T-wave signal. The optical power LO $P_{LO}(t)$ has an amplitude $P_{LO,0} = \hat{P}_{LO,1} = 80\,\text{mW}$ and oscillates with frequency $f_{LO} = 0.31\,\text{THz}$. The noise power measured at the IF output of the Rx module drops significantly for increasing frequencies with a similar characteristic as the Rx response shown in Fig. S3 such that the signal-to-noise power ratio remains essentially constant. The origin of the strong frequency dependence of the Rx response and the Rx noise needs further investigation.

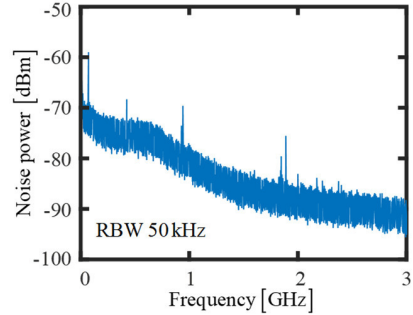

**Fig. S4.** Noise power measured at the output of the Rx without an incident T-wave. The LO optical power $P_{LO}(t)$ has an amplitude $P_{LO,0} = \hat{P}_{LO,1} = 80\,\text{mW}$ and oscillates with frequency $f_{LO} = 0.31\,\text{THz}$.

## 3. Wireless THz communication link

The detailed experimental setup used for the communication experiments is shown in Fig. S5(a). Tunable laser sources with linewidth smaller than 100 kHz (Keysight, N7714A) provide the optical tones for the Tx and the Rx. At the Tx, an arbitrary-waveform generator (AWG) is used to drive an IQ-modulator which is fed by an optical c.w. carrier with frequency $f_{S,a}$. As a data signal, we use a De Bruijn sequence[9] of length $2^{13}$. The optical signal is amplified by an erbium-doped fiber amplifier (EDFA), followed by a 0.6 nm filter to suppress amplified spontaneous emission (ASE) noise. A 50/50-coupler combines the modulated carrier with an unmodulated c.w. tone at frequency $f_{S,b}$. To ensure strong interference of the two optical signals, we adjust the polarization using two polarization controllers (Pol. Contr.) by maximizing the power after a polarizer (Pol.). An optical attenuator (Keysight N7764A) is used to set the power level $P_S(t)$. Finally, we adjust the polarization of the combined signal to maximize the current in a high-speed uni-travelling-carrier photodiode (UTC-PD). In the UTC-PD, the optical signal is converted to a T-wave signal with frequency $f_S = |f_{S,a} - f_{S,b}|$ (O/T - conversion). The T-wave is radiated to free space by a horn antenna and a PTFE lens (Thorlabs, LAT200).

After a transmission distance of 58 m, the T-wave is captured by another PTFE lens and coupled to a WR 3.4 hollow waveguide by a horn antenna. The waveguide is connected to the input of two cascaded T-wave amplifiers[10], which compensate the free-space transmission loss and amplify the T-wave. In our current design, we use another horn antenna at the output of the second T-wave amplifier in combination with a silicon lens to couple the T-wave to the photoconductor. In the future, the performance of the scheme may be further improved by replacing this assembly with a waveguide-coupled photoconductor. For generating the photonic LO, two c.w. laser tones with optical frequencies $f_{LO,a}$ and $f_{LO,b}$ are superimposed using a polarization-maintaining 50/50 coupler, thus generating an optical power beat. The beat signal is amplified by an EDFA followed by a 3 nm filter to reduce ASE noise. A polarization controller is used to maximize the IF signal at the output of the polarization-sensitive photoconductor. With an attenuator we adjust the optical power $P_{LO}$ at the Rx. Finally, the down-converted IF signal is coupled to the TIA, the output voltage of which is sampled and stored in a real-time oscilloscope for further offline signal processing. Figure S5(b) shows a photograph of the wireless transmission link. The image on the left shows the Tx including the UTC-PD and the T-wave PFTE lens. The Rx is 58 m away from the Tx. On the right image, the Rx including the T-wave PTFE lens, the T-wave amplifiers and the Rx module is shown in more detail. To facilitate identification of the components shown in the setup sketch of Fig. S5(a), we mark them with the letters A - H.

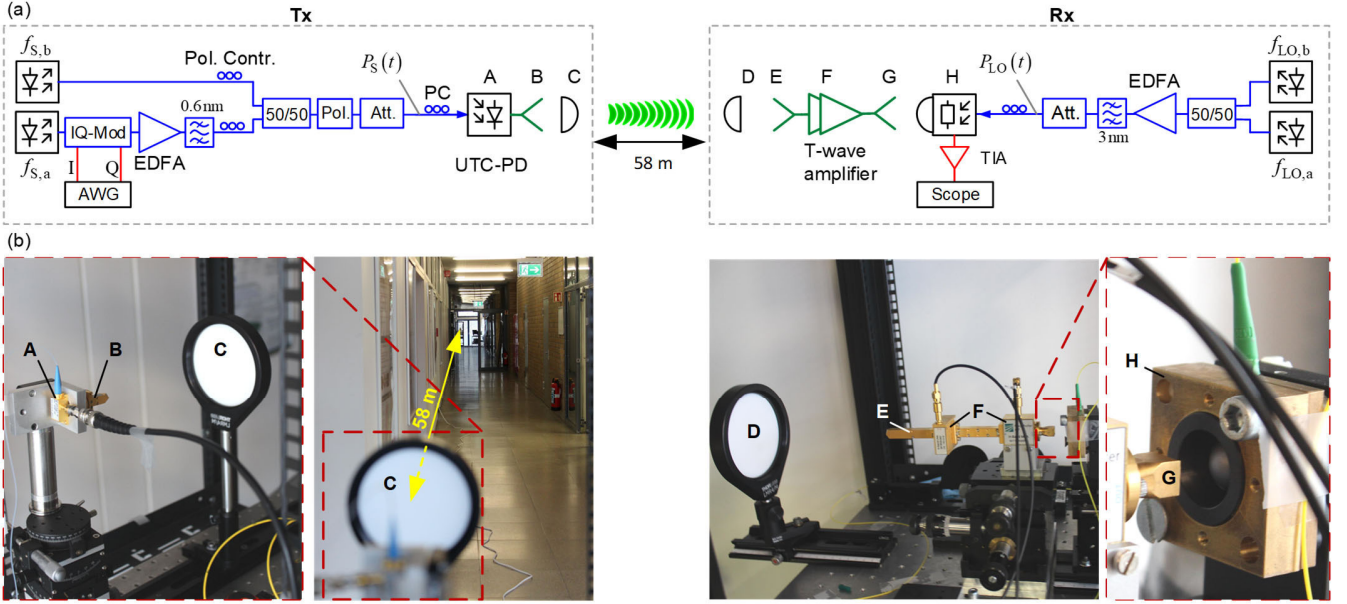

**Fig. S5.** Details of wireless data transmission link. **(a)** Schematic of the T-wave Tx and Rx used in our experiments. **(b)** Photograph of Tx and Rx. Components in the schematic of subfigure (a) are marked with letters A…H. At the Tx, the UTC-PD (A), the horn antenna (B), and the PTFE lens (C) are shown. A distance of 58 m is bridged between the Tx and Rx. The Rx setup contains a second PTFE lens (D), the amplification stage including an input horn antenna (E), the T-wave amplifiers (F), the output horn antenna (G), and the T-wave Rx module (H).

For finding optimum operation parameters, we characterize the performance of the wireless link shown in Fig. S5 for different optical powers $P_S(t)$ and powers $P_{LO}(t)$ at the Tx and the Rx,

$$P_S(t) = P_{S,0} + \hat{P}_{S,1}\cos(\omega_S t), \quad f_S = |f_{S,a} - f_{S,b}|,$$
$$P_{LO}(t) = P_{LO,0} + \hat{P}_{LO,1}\cos(\omega_{LO} t), f_{LO} = |f_{LO,a} - f_{LO,b}|. \quad (S9)$$

In our measurements, we adjust the lasers at the Tx and Rx such that the average powers $P_{S,0}$ and $P_{LO,0}$ are equal to the respective oscillation amplitudes, $P_{S,0} = \hat{P}_{S,1}$ and $P_{LO,0} = \hat{P}_{LO,1}$. The Tx and Rx frequencies are set to $f_S = 0.310\,\text{THz}$ and $f_{LO} = 0.309\,\text{THz}$, respectively. For studying the transmission performance, we generate a quadrature phase-shift keying (QPSK) signal with a line rate of $R_b = 2\,\text{Gbit/s}$ at the Tx and measure its BER after down-conversion at the Rx.

Figure S6(a) shows the BER (red dots) obtained for various optical Tx powers $P_{S,0} = \hat{P}_{S,1}$ and for a constant optical LO power of $P_{LO,0} = \hat{P}_{LO,1} = 80\,\text{mW}$ at the Rx. For some measurement points, the signal quality is so high that we could not measure any errors in a recording length of $10^5$ symbols. We therefore estimate the BER from the error vector magnitude[11] (blue dots). For an optical power of $\hat{P}_{S,1} > 8\,\text{mW}$, the signal quality decreases because the T-wave amplifiers saturate, see constellation diagrams in the right-hand side column of Fig. S6(a). For high optical powers $\hat{P}_{S,1}$, saturation of the T-wave amplifiers leads to an asymmetric distribution of the noise around the various constellation points, whereas a symmetric distribution is observed for low optical powers.

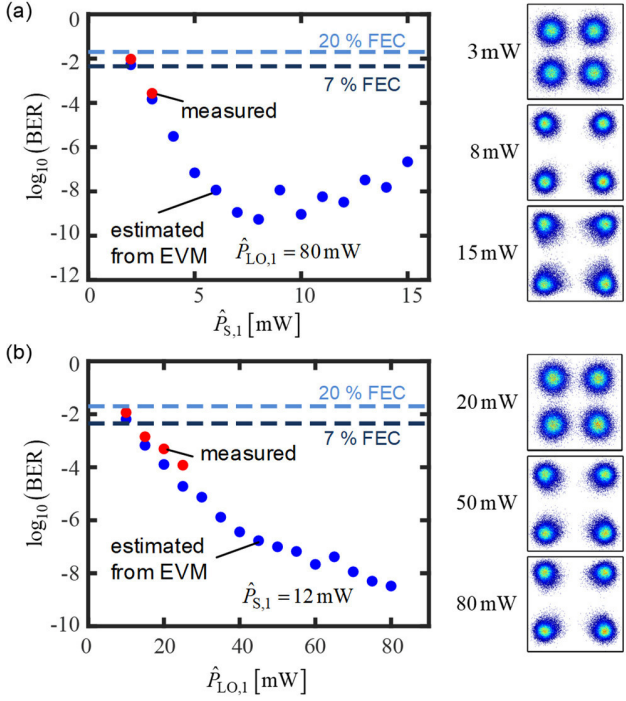

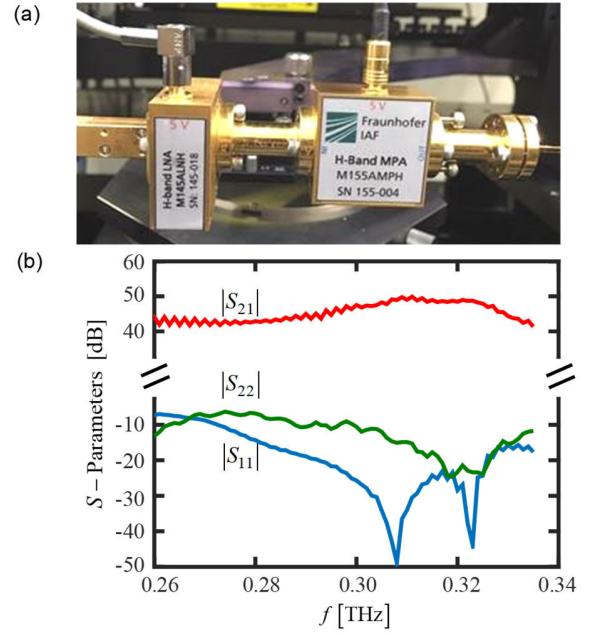

**Fig. S6.** Measured and estimated BER for different optical Tx and LO powers. As a test signal, we transmit QPSK data with a line rate of $R_b = 2\,\text{Gbit/s}$. The T-wave carrier frequency is set to $f_S = 0.310\,\text{THz}$ **(a)** BER as a function of the Tx power amplitude $\hat{P}_{S,1}$. Red dots denote values that were directly measured, whereas blue dots refer to BER values estimated from the respective error vector magnitude (EVM). Since the length of our signal recordings was limited to $10^5$ symbols, the lowest statistically reliable measured BER amounts to $10^{-4}$. For measured BER values above this threshold, directly measured and estimated BER show good agreement, giving us confidence that the EVM-based estimations for BER $< 10^{-4}$ are valid. **(b)** BER as a function of the LO power amplitude $\hat{P}_{LO,1}$.

Figure S6(b) shows the BER measured for various optical LO powers $P_{LO,0} = \hat{P}_{LO,1}$ at the Rx. In this case, the optical Tx power is kept constant at $P_{S,0} = \hat{P}_{S,1} = 12\,\text{mW}$, close to its optimum point shown in Fig. S6(a). The signal quality improves with increasing optical LO power $\hat{P}_{LO,1}$ and is finally limited by the maximum optical power that the photoconductor can withstand. 4. T-wave amplifiers and UTC-PD

To compensate free space T-wave transmission loss, we use a cascade of a low-noise amplifier (LNA) and a medium-power amplifier (MPA), designed for operation in the submillimeter H-band (0.220 THz - 0.325 THz), see Fig. S7(a). The moduli of the *S*-parameters for this cascade are shown in Fig. S7(b). In a frequency range from 0.260 THz to 0.335 THz, the total gain is more than 40 dB.

**Fig. S7.** Characterization of cascaded T-wave amplifiers. **(a)** Photograph of low-noise amplifier (LNA) and medium paper amplifier (MPA). **(b)** *S*-parameters measured by a vector network analyzer with frequency extension modules. A total gain of over 40 dB is achieved in a frequency range from 0.260 THz to 0.335 THz.

To measure the frequency response of T-wave components and of the complete transmission system, we use the setups shown in Fig. S8. Two unmodulated c.w. laser tones having equal powers and different frequencies $f_{S,a}$ and $f_{S,b}$ are superimposed in a 50/50 combiner and coupled to the UTC-PD. The T-wave output power $P_{THz}$ is measured in a calorimeter (VDI, Erickson PM4). By tuning the difference frequency $f_S = |f_{S,a} - f_{S,b}|$ of the two lasers we measured the frequency dependent output power of the UTC-PD without any amplifier, with the MPA, or with the cascade of LNA and MPA, Fig. S8(a,b,c). Furthermore, we measured the power after T-wave transmission over 58 m with the cascaded LNA and MPA at the Rx, Fig. S8(d). The results of all these measurements are shown in Fig. 8(e,f). Note that optical input power of the UTC-PD had to be strongly reduced for the case of the cascaded T-wave amplifiers without free-space link to prevent amplifier saturation. The received THz power after the amplifier cascade is more than 0 dBm in a frequency range from 0.29 THz to 0.33 THz for a transmission distance of 58 m.

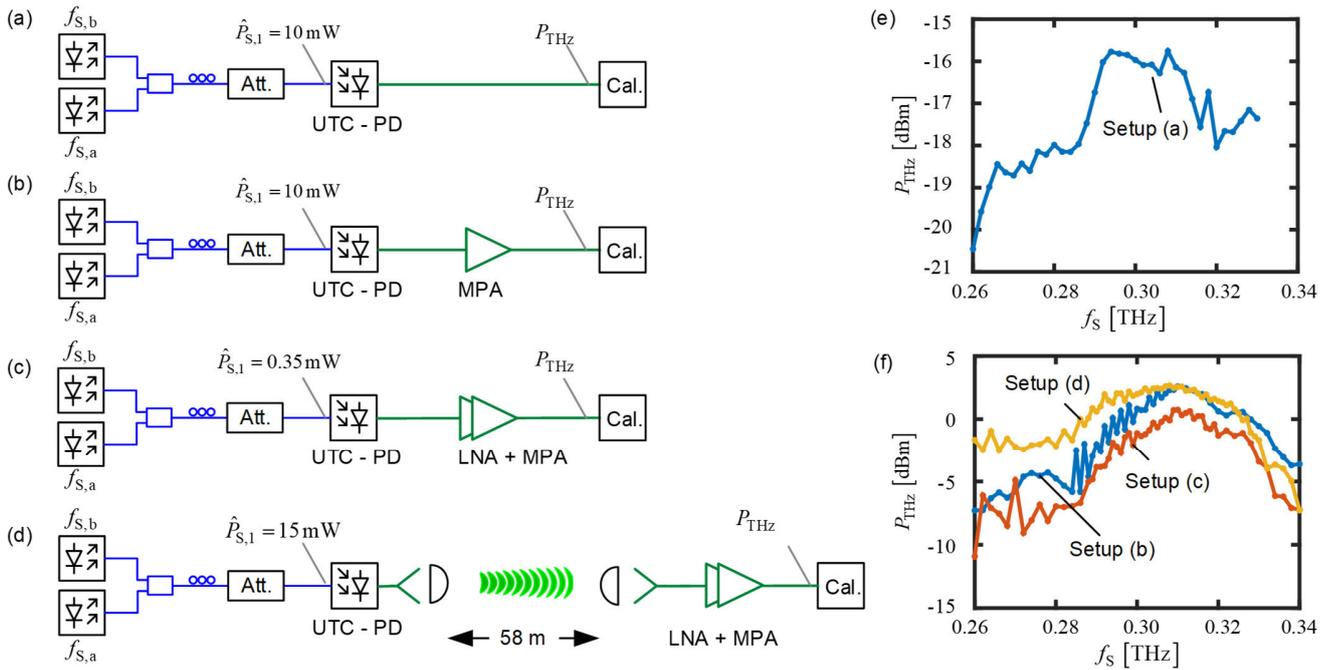

**Fig. S8.** Setup and results for measuring T-wave output power of UTC-PD and T-wave amplifiers. **(a)** Setup to measure T-wave output power $P_{\text{THz}}$ of UTC-PD as a function of frequency $f_S = |f_{S,a} - f_{S,b}|$. **(b)** T-wave output power of UTC-PD and MPA. **(c)** T-wave output power of UTC-PD, LNA and MPA. **(d)** T-wave output power of UTC-PD, 58 m free space transmission, and LNA-MPA cascade. **(e)** Measured T-wave output power $P_{\text{THz}}$ in dependence of Tx frequency $f_S$ for UTC-PD. **(f)** Measured T-wave output power $P_{\text{THz}}$ for UTC-PD and MPA, setup Fig. S8(b), for UTC-PD and LNA-MPA cascade, setup Fig. S8(c), and for UTC-PD, 58 m free space transmission and LNA-MPA cascade, setup Fig. S8(d).